\documentclass[article,aps,amsmath,amssymb,letterpaper,twocolumn]{revtex4}

\pdfoutput=1
\usepackage{hyperref}
\usepackage{graphicx}
\def\cal#1{\mathcal{#1}}
\def\eq#1{Equation~(\ref{#1})}

\def\beq{\begin{equation}}
\def\eeq{\end{equation}}
\def\bea{\begin{eqnarray}}
\def\eea{\end{eqnarray}}
\usepackage{bm}

\begin{document}
\title{Transformation from spots to waves in a model of actin pattern formation}
\author{Stephen Whitelam\email{swhitelam@lbl.gov}$^{1,2}$, Till Bretschneider$^{2}$ and Nigel J. Burroughs$^{2}$} 
\affiliation{$^{1}$Molecular Foundry, Lawrence Berkeley National Laboratory, 1 Cyclotron Road, Berkeley, CA 94720, USA\\
$^{2}$Systems Biology Centre, University of Warwick, Coventry, CV4 7AL, UK}
\begin{abstract}
Actin networks in certain single-celled organisms exhibit a complex pattern-forming dynamics that starts with the appearance of static spots of actin on the cell cortex. Spots soon become mobile, executing persistent random walks, and eventually give rise to traveling waves of actin. Here we describe a possible physical mechanism for this distinctive set of dynamic transformations, by equipping an excitable reaction-diffusion model with a field describing the spatial orientation of its chief constituent (which we consider to be actin). The interplay of anisotropic actin growth and spatial inhibition drives a transformation at fixed parameter values from static spots to moving spots to waves.
%Actin networks in certain single-celled organisms exhibit a complex pattern-forming dynamics that starts with the appearance of static `spots' of actin on the cell cortex. Spots soon become mobile, executing persistent random walks, and eventually give rise to and coexist with traveling waves of actin. Waves confer motility upon the cell as they impinge on its periphery. Here we describe a possible physical mechanism for this distinctive set of dynamic transformations. Starting from the observation that excitable reaction-diffusion models of chemical systems can describe both localized spots and traveling waves, we augment one such model with a variable describing the spatial orientation of its chief constituent. We consider this constituent to be a caricature of polymerized actin, and regard its orientation as a measure of local actin fiber alignment. The bias exerted on network growth by this anisotropy profoundly affects localized structures to drive a transformation at fixed parameter values from static spots to moving spots to waves. 
\end{abstract}
\maketitle
\begin{figure}[h!] 
\label{}
\centering
\includegraphics[width=7cm]{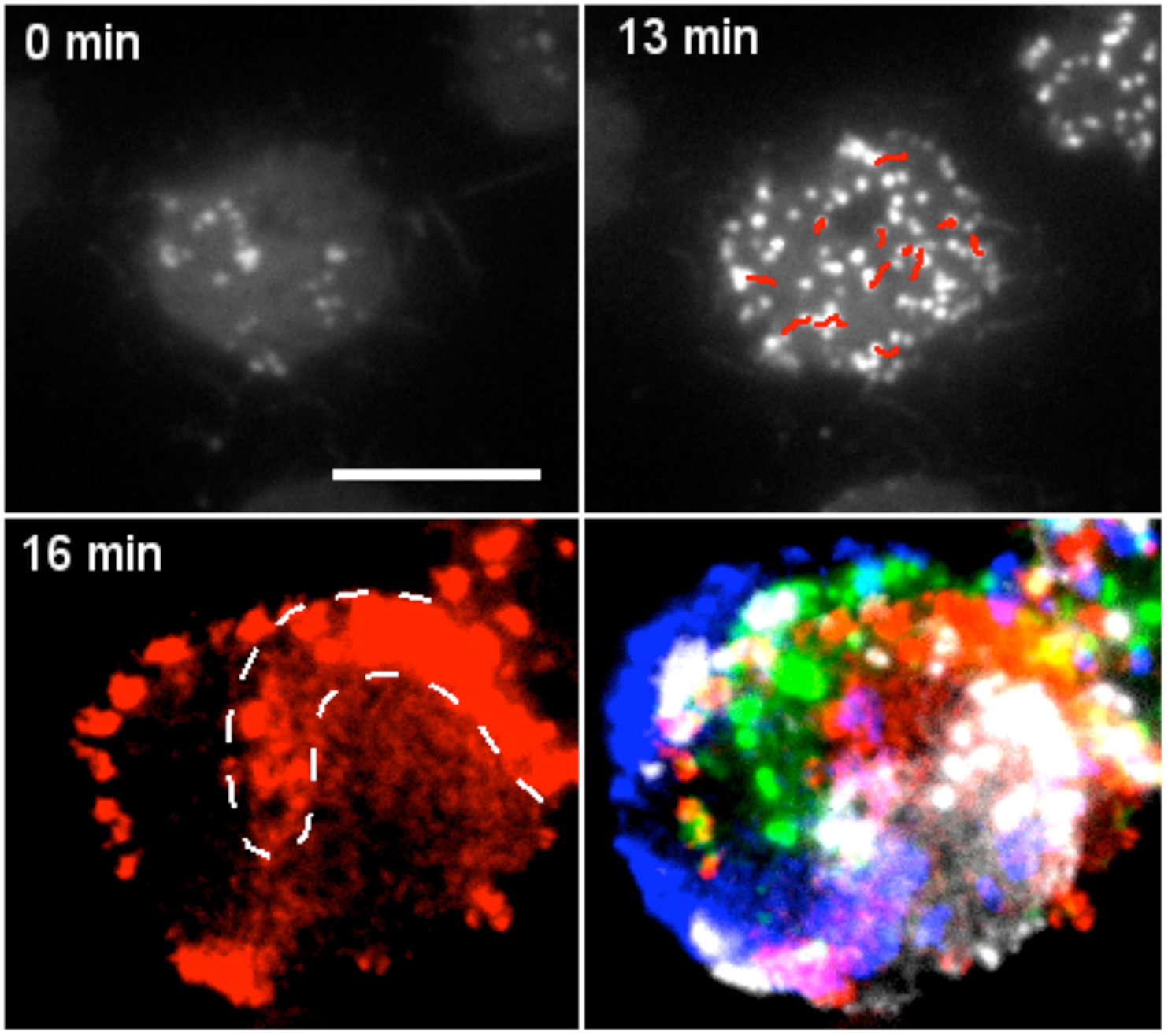} 
\caption{\label{figtill} Dynamic actin structures in the substrate-attached cortex of a {\em Dictyostelium} cell, as visualized by TIRF microscopy. LimE$\Delta$coil-GFP was used to selectively label F-actin. The cell was incubated for 15 min under 5 $\mu$M latrunculin A to depolymerize F-actin. The first frame (0 min) is taken 16 minutes after reducing latrunculin A concentration to 1 $\mu$M; the first stationary actin spots are visible (scale bar 10 $\mu$m). Spots become more numerous and mobile (13 min frame; selected traces over a period of 30 sec shown in red). At 16 min a prominent spiral wave appears (see dashed outline). Bottom right: 4 frames of the counterclockwise-rotating spiral wave superimposed in different colors (red, green, blue, white; first and final images separated by 45 sec).}
 \end{figure}

{\em Introduction.} {\em Dictyostelium discoideum} ({\em Dicty}) is an amoeba known to generate spectacular patterns through organized multicellular aggregation~\cite{gerisch1968caa}. This organization is made possible by {\em Dicty}'s ability to move, which in turn is regulated by polymerization of the protein actin into oriented networks within individual amoebae~\cite{pollard2003ccm}. Interestingly, these networks exhibit their own distinctive patterns. When treated with the drug latrunculin, actin networks within {\em Dicty} degrade, rendering the amoeba immobile. Upon removal of latrunculin, actin networks re-polymerize through a complex pattern-formation process that appears to consist of three stages~\cite{gerisch2004mac,till_new}, summarized in Fig.~\ref{figtill}. First, immobile circular spots of actin form on the cell membrane. Second, spots acquire a persistent diffusive motion. Third, wave-like actin structures reminiscent of cell `leading edges' appear and coexist with spots. As waves strike the cell periphery the amoeba recovers its ability to move.

This paper describes a mechanism that might underpin the remarkable dynamic transformation from static spots to moving spots to waves. Our starting point is the recognition that spots (both static and moving) and waves have been seen in chemical systems~\cite{cross1993pfo}, and that such patterns can be described by equations modeling reacting and diffusing chemicals~\cite{vanag2007lpr}. Here we consider an excitable `activator-inhibitor' reaction-diffusion model that describes, in different regions of its parameter space, stationary spots and moving waves. In order to interpret the chief constituent of this model (the `activator') as substrate-bound actin in {\em Dictyostelium}, we equip it with an additional variable that describes local actin fiber orientation. We find that the resulting model exhibits a series of nonequilibrium transformations, at fixed parameter values, from stationary spots to moving spots to traveling waves. This transformation is driven by the interplay of inhibition (which permits localized spots) and the directional bias imparted to actin polymerization by local fiber orientation, the latter emerging from a spontaneous breaking of fiber symmetry. Our results support the idea that emergent actin patterns {\em in vivo} are fostered by an excitable medium~\cite{vicker2000rdw,weiner2007abw}, and illustrate the dynamical richness accessible to an intrinsically anisotropic chemical species that suffers inhibition. Models of inhibited but spatially {\em isotropic} chemicals can describe striking transitions between static and moving spots, but only as their parameters are varied~\cite{krischer1994bts}; traditional models of anisotropic actin polymerization (treadmilling) do not possess as solutions static, size-limited, isolated spots.
 
We present our model below. We discuss the origin of spot stability, and demonstrate numerically that local fiber orientation can both induce a spot to move and distort a spot into a wave. The resulting dynamical evolution of the model captures several features of the spots and waves seen {\em in vivo}.

{\em Model.} The FitzHugh-Nagumo equations~\cite{fitzhugh1969mme,nagumo} studied by Vasiev~\cite{vasiev} describe localized spots and waves in different regions of parameter space. We take these equations as our starting point in an attempt to describe similar patterns seen in {\em Dicty}, and augment these equations to account for the polarity of the fibers comprising polymerized actin structures. We consider the evolution of scalar fields $u({\bm x},t)$ and $v({\bm x},t)$ on a two-dimensional substrate (cell membrane) according to the equations
\bea
\label{fitzhugh1}
\partial_t u({\bm x},t)&=& \nabla \cdot {\bm J} +h(u)-\rho v+ \sqrt{2T_u} \eta({\bm x},t); \\
\label{fitzhugh2}
\partial_t v({\bm x},t)&=& \delta \nabla^2 v+\epsilon(u-v).
\eea
Here ${\bm J} \equiv \nabla u + V_0 {\bm \tau} u$ and $h(u) \equiv -k u(u-u_1)(u-u_0)$. We shall set the parameters $\epsilon$, $\delta$, $\rho$, $k$, $u_0$ and $u_1$ by comparison with previous work~\cite{vasiev}, and will explore the effect of varying $V_0$ and $T_u$. We consider $u({\bm x},t)$ to be proportional to the number of actin fibers per unit area (relative to a reference concentration) at substrate position ${\bm x}$ at time $t$. The field $v({\bm x},t)$ acts to degrade actin, and the vector field ${\bm \tau}({\bm x},t)$ labels the orientation of the actin network. Equations~(\ref{fitzhugh1}) and (\ref{fitzhugh2}) (with $V_0=T_u=0$) describe classical excitable behavior wherein the autocatalytic `activator' field $u$ and activator-suppressing `inhibitor' field $v$ interact to generate localized patterns that may propagate spatially~\cite{murray2003mb}. We consider the homogeneous terms of \eq{fitzhugh1} to describe polymerization of actin at a rate proportional its local concentration,
%~\footnote{The quadratic dependence of this rate on local actin concentration assumes a polymerization-enhancing feedback mechanism such as that which might operate when cortex-bound myosin I recruits actin, which in turn recruits more myosin.}, 
$r_{\rm pol} = k(u_1+u_0) u^2$, and degredation of actin with rate $r_{\rm deg} = -k u_0 u_1 u$.  We regard the term $-k u^3$ as a model of the concentration-limiting effect of steric hinderance. The field $\eta$ is a Gaussian white noise with zero mean and unit variance; $T_u \geq 0$ quantifies the magnitude of this noise. 

\begin{figure}[h!] 
\label{}
\centering
\includegraphics[width=8cm]{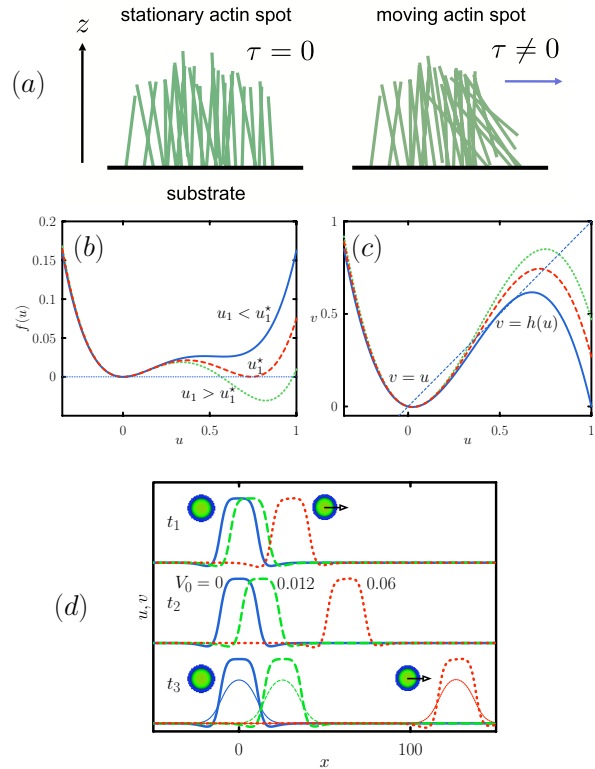} 
\caption{\label{comp1} Stationary and moving model spots. (a) Cartoon of the actin fiber structure we conjecture for stationary and moving spots {\em in vivo}. (b) Homogeneous effective `free energy' density $f(u)$ from our model (with $u_0=0.05,k=4.5$~\cite{vasiev}) for values of actin polymerization rate $u_1$ below ($u_1=1$~\cite{vasiev}; solid line) at ($u_1=u_1^{\star}\approx 1.06$; dashed line) and above ($u_1=1.1$; dotted line) its first-order critical point; (c) nullclines of Equations (\ref{fitzhugh1}) and (\ref{fitzhugh2}) under the same conditions. (d) $u$-field profiles (1$d$ cuts through a 2$d$ box of size $500^2$) at three different times ($(t_1,t_2,t_3)=(5,10,20)\times10^3$) from three different simulations $(V_0=0,0.012,0.06)$. We excite the medium at $t=0$ so that a spot nucleates, and we impose an orientation field ${\bm \tau}= \hat{ {\bm x}}/2$. The spot is stationary and stable for $V_0=0$; the spot moves persistently without significant structural distortion at the two larger values of $V_0$. We show at $t_3$ the $v$-field profiles also (thin lines). Parameter values: $u_0=0.05,u_1=1,k=4.5,\epsilon=0.5,\delta=2.5,T_u=0$.}
 \end{figure}

We motivate the terms in \eq{fitzhugh1} describing spatial propagation of actin in the following way. We hypothesize (on the basis of  bleaching experiments in {\em Dicty}~\cite{till_new}) that stationary spots are composed of fibers oriented principally normal to the substrate (Fig.~\ref{comp1}(a)). We assume that membrane-bound proteins (such as MyoB) recruit Arp2/3 complexes to the growing ends of fibers, and that these complexes in turn initiate the growth of daughter fibers at the spot periphery. We argue that such growth is isotropic, and the resulting propagation of material diffusive. By contrast, we expect that waves possess many fibers aligned in part parallel to the substrate and move chiefly by treadmilling~\cite{pollard2003ccm,kruse}. To describe fiber orientation we have introduced a field ${\bm \tau}({\bm x},t)$: vanishing ${\bm \tau}$ describes fibers pointing solely in the vertical direction, while nonzero ${\bm \tau}$ describes fibers with some component of orientation parallel to the substrate. We assume that lateral fiber orientation biases the direction of actin growth, which we model using the term $V_0 \nabla \cdot  {\bm  \tau} u $. $V_0$ controls the rate of directed polymerization. We further hypothesize that orientation may be acquired spontaneously, and therefore require that ${\bm \tau}$ evolve according to the equation
\beq
\label{three}
\Gamma_{\tau}^{-1} \partial_t {\bm \tau}({\bm x},t) =-\frac{\delta}{\delta \bm  \tau} \int d^2 {\bm x} \, {\cal F}_{ \tau}[{\bm \tau},u]+ \sqrt{2T_{\tau}} \, \xi({\bm x},t).
\eeq
Here ${\cal F}_{\tau} \equiv h_0\tau  - u \tau^2 +\tau^4+ \alpha_1 \left( {\bm \nabla} \cdot {\bm \tau} \right)^2+  \alpha_2 \left( {\bm \nabla} \times {\bm \tau} \right)^2$ is a Landau-esque free energy density describing a field ${\bm \tau}$ that may order in the presence of the field $u$ (provided that the `barrier' to initial ordering, the term in $h_0$, is sufficiently small) and that has a tendency to align. Its degree of alignment is determined by a competition between the orientation-inducing terms coupled to $\alpha_{1,2}$, and the orientation-destroying effect of the noise term coupled to $T_{\tau}$. The field $\xi$ is a Gaussian white noise with zero mean and unit variance. We set (arbitrarily) $\alpha_1=\alpha_2=0.2$, and regard $T_{\tau}$ as our chief measure of network-alignment propensity. The kinetic prefactor $\Gamma_{\tau}$ controls the rate of network alignment relative to that of network polymerization.
 
\eq{fitzhugh2} describes the evolution of an actin-suppressing `inhibitor field'. Actin recruits the agent of its destruction at rate $\epsilon u$; this agent degrades actin at a rate $-\rho v$, and is itself degraded at a rate $-\epsilon v$. Possible biological origins for $v$ include the concentration of actin severing proteins (e.g. cofilin) and actin capping proteins, or the state of hydrolysis of fibers. Here we simply regard $v$ as a coarse-grained agent of actin degredation. Our model assumes that actin polymerization is not in thermal equilibrium~\cite{pollard2003ccm}. 
%(hydrolyzed actin degrades more rapidly than does ATP-bound actin). The term $\delta \nabla^2 v$ would then quantify either protein molecular diffusion or accord with the `vectorial hydrolysis' mechanism in which regions adjacent to hydrolyzed actin are more likely to hydrolyze than are regions far from hydrolyzed actin~\cite{carlier}. 

{\em  Intuitive explanation of spot stability.} Other authors have demonstrated semi-numerically~\cite{vasiev} and analytically~\cite{ohta,petrich1994ncd,goldstein1996ipa} that equations similar to (1) and (2) with $V_0=0$ admit static spots as steady-state solutions in certain parameter regimes (a stationary spot profile is shown in Fig.~\ref{comp1} (d)). Here we follow approaches detailed in Refs.~\cite{petrich1994ncd, goldstein1996ipa,krischer1994bts,gompper1992glt} to put forward a simple physical argument for why such spots can exist. At steady state, and with coordinates rescaled such that $\nabla^2 \to (\epsilon/\delta) \nabla^2$, \eq{fitzhugh2} has solution $v({\bm x}) = (2 \pi)^{-1} \int d {\bm x}'    K_0(| {\bm x}'-{\bm x}|) u({\bm x'})$.  We expand this solution to second order in $\nabla$ and insert the resulting expression in the steady-state version of \eq{fitzhugh1}. We regard the equation so obtained as the Euler-Lagrange equation of the `free energy' functional
 \beq
 \label{func}
 F_u[u] = \int d^2 {\bm x} \left\{ f(u)+ g_{\rho} \left( \nabla u\right)^2 + c_{\rho} \left(\nabla^2 u \right)^2 \right\},
 \eeq
where $f(u) \equiv a_4 u^4-a_3 u^3+a_2(\rho) u^2$; $a_4 \equiv k/4$; $a_3 \equiv  k (u_0+u_1)/3$; $a_2(\rho) \equiv (k u_1 u_0+\rho)/2$; $g_{\rho} \equiv  \left(\epsilon/\delta - \rho \right)/2$; and $c_{\rho} \equiv \rho/2$. The homogeneous component $f(u)$ describes a first-order phase transition from an empty substrate to an actin-covered substrate as the parameter $u_1$ is increased ($u_1$ is roughly proportional to the mean-field actin polymerization rate); bulk phase coexistence occurs when $u_1=u_1^{\star} = \frac{5}{4} u_0 +\frac{3}{4} \left(u_0^2 + 8\rho/k\right)^{1/2}$. For the parameters of reference~\cite{vasiev} ($\rho=1$, $u_0=0.05$ and $k=4.5$), we have $u_1^{\star} =1.0632$; in that paper, $u_1$ was set to unity, implying that the system considered there was numerically close to coexistence. We show $f(u)$ (and its corresponding nullclines) in this regime in Fig.~\ref{comp1}(b) (Fig.~\ref{comp1}(c)). Note that the activator-inhibitor coupling $\rho$ influences the location of the mean-field critical point, but not the nature of the transition.

The space-dependent terms of~\eq{func} admit modulated patterns when $\rho>0$ and $\epsilon/\delta<\rho$; in this paper we perform simulations with $\rho=1,\epsilon=0.5$ and $\delta=2.5$. It is instructive to recognize that while versions of \eq{func} appear in many different settings~\cite{goldstein1996ipa}, the inhibitor-induced spatial modulation of the $u$ field closely resembles the physical mechanism by which surfactant induces microphase separation of oil in water~\cite{gompper1992glt}. One might therefore consider the stable spots of the FitzHugh-Nagumo equations studied by Vasiev~\cite{vasiev} to be akin to drops of surfactant-coated oil in water. Moreover, extremizing~\eq{func} in circular geometry for an assumed density profile (adapting the approach of Ref.~\cite{gompper1992glt}) reveals a characteristic spot size. While a local approximation such as~\eq{func} possesses limited predictive power~\cite{goldstein1996ipa}, it does capture the correct trend of variation of spot radius with model parameters (such as $\epsilon$), and provides an intuitive explanation for why static spots of well-defined size exist in this system. 
\begin{figure}[ht]
\label{}
\centering
\includegraphics[width=9cm]{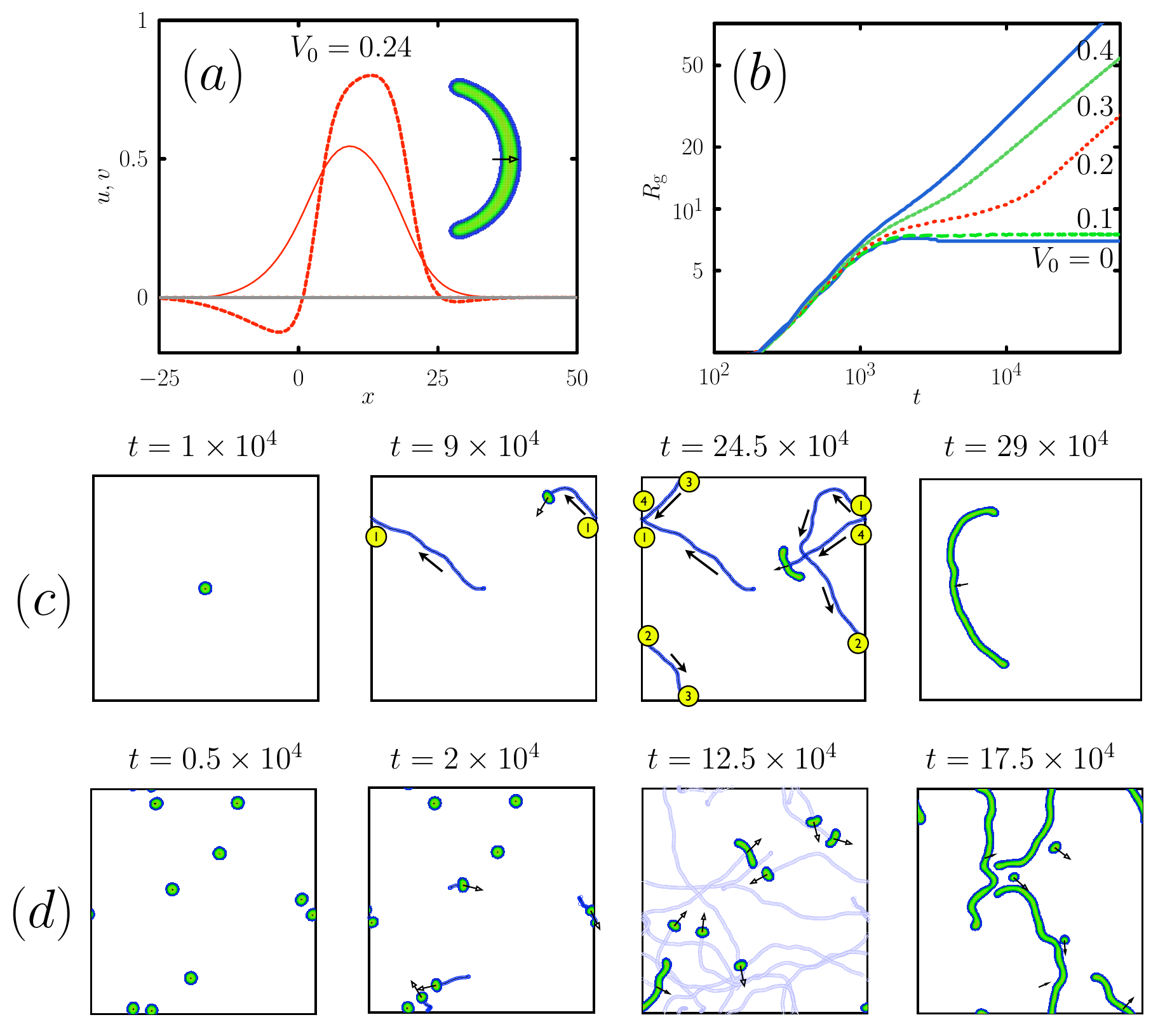} 
\caption{\label{fig3} Model spot-to-wave transformation. (a) An initially static spot moves under the influence of a fixed orientation field ${\bm \tau}=\hat{{\bm x}}/2$ $(V_0=0.24)$ and deforms into a structure resembling a traveling wave (profile and image shown $1.05 \times 10^5$ time units after application of the field). (b) We quantify this distortion mechanism by plotting the radius of gyration $R_{\rm g}$ as a function of time for spots exposed to similar fixed ${\bm \tau}$ and various values of $V_0$. (c) Time-ordered snapshots of $u(\mathbf{x})$ from a simulation with fluctuating ${\bm \tau}$ starting from a single spot in the center of a box. The spot acquires orientation spontaneously (spot arrow labels spot-averaged ${\bm \tau}$), executes a persistent random walk (numbers identify periodic box crossings; trajectories and arrows record spot motion), and begins to deforms into a wave at $t\approx 24 \times 10^4$. (d) A similar simulation with nine initial spots mimics the transformation from static spots to moving spots to waves shown in Fig.~\ref{figtill}.}
 \end{figure}
 
{\em Effect of fiber orientation upon spots}. To determine the effect of ${\bm \tau}$ upon the behavior of spots, we performed numerical simulations of our model for a parameter set that admits, at steady state, immobile spot solutions when $V_0=0$ ($u_0=0.05,u_1=1,k=4.5,\epsilon=0.5,\delta=2.5$~\cite{vasiev}). We integrated Equations (1--3) using the second-order Heun algorithm~\cite{greiner1988nis} on a square grid of lattice constant 0.4 with timestep 0.01~\cite{vasiev}. We used a 9-point Laplacian~\cite{ninepoint}. In Fig. \ref{comp1}(d) we show that when ${\bm \tau}$ and $V_0$ are fixed, a spot moves persistently. Moving spots in our model behave similarly to the moving spots described in Ref.~\cite{krischer1994bts}, colliding both inelastically (at high speeds) and elastically (at low speeds). When $V_0$ is made sufficiently large, a spot deforms into a localized structure (Fig. \ref{fig3}(a,b)) resembling the traveling waves seen in the regular FitzHugh-Nagumo model at parameter values distinct from those at which spots are found~\cite{vasiev}. This exploitation of a spot's apparent underlying instability to wave deformation, which embodies the notion of nonequilibrium pattern control discussed in Ref.~\cite{hagberg1996cdp}, requires persistence of orientation. In the full model, when ${\bm \tau}$ evolves according to \eq{three} (with noise), the transformations from static spot to moving spot and from moving spot to wave can occur on a broad range of timescales. In Fig. \ref{fig3}(c) we show configurations of $u({\bm x})$ from a simulation $(\Gamma_u=1,h_0=0.225,T_u=0, T_{\tau}=0.015,V_0=0.1)$ that starts from a single spot at the center of a box of size 450$^2$, periodically replicated in imitation of bulk surroundings. The spot spontaneously acquires orientation (at $t \approx 1.3 \times 10^4$), executes a persistent random walk, and eventually deforms into a wave. Spot trajectories are shown in the second and third panels, with periodic boundary crossings labeled in the order they occur. In Fig. \ref{fig3}(d) we show images from a similar simulation with nine initial spots. A phase of autonomous mobile spots is supplanted at later times by an organized collection of waves (moving from left to right), mimicking the transformation seen in {\em Dicty}. Indeed, the model's dynamics captures several qualitative features seen {\em in vivo}~\cite{gerisch1968caa,till_new}: spots nucleate spontaneously (when $T_u \gtrsim 0.05$); spots spontaneously (and autonomously) acquire orientation; oriented spots execute persistent random walks; spots split (cf.~\cite{wang2007scs}) and merge; moving spots deform into larger, wave-like structures on a broad range of deformation times; waves and spots may coexist; and large structures degrade or limit the growth of neighboring smaller structures. 

We have shown that a simple and physically-motivated modification of the FitzHugh-Nagumo equations can mimic the interconversion and coexistence of distinct dynamical motifs seen in {\em Dictyostelium}. Spatially localized patterns in reaction-diffusion systems have recently received considerable attention~\cite{vanag2007lpr}, and the model we present demonstrates the unusually rich dynamics accessible to an intrinsically {\em anisotropic} chemical species. Our work suggests that certain cytoskeletal dynamics can indeed be caricatured by simple reaction-diffusion models embodying the idea of excitability~\cite{vicker2000rdw,weiner2007abw}, supporting a physical picture (that one might dub the {\em excitoskeleton}) in which cytoskeletal pattern formation is driven by the self-organization of excitable solitons of actin and its attendant proteins, rather than being orchestrated solely by biophysical signaling pathways. 
 
We thank B.N. Vasiev for discussions. Work at the Molecular Foundry was supported by the U.S. Department of Energy under Contract No. DE-AC02-05CH11231. Support at Warwick was provided by the BioSim E.U. Network of Excellence and Warwick's Centre for Scientific Computing.
 
\bibliography{bib}

\end{document}